\documentclass[a4paper,11pt]{article}

\usepackage{pos}
\allowdisplaybreaks[4]                   
\newcommand{\ep}{\varepsilon}

\title{{\footnotesize
DESY 19--061, DO-TH 20/09, TTP22--048, 
RISC Report Series 22--09, SAGEX 20--20, PoS (LL2022) 048}\\
The 3-loop anomalous dimensions from off-shell operator 
matrix elements\footnote{
This project has received funding from the European Union’s Horizon 2020 research and innovation
programme under the Marie Sklodowska–Curie grant agreement No. 764850, SAGEX and from
the Austrian Science Fund (FWF) grants SFB F50 (F5009-N15) and P33530.}}
 \ShortTitle{3-loop anomalous dimensions}

\author*[a]{Johannes Bl\"umlein}
\author[a]{Peter Marquard}
\author[b]{Carsten Schneider}
\author[c]{Kay Sch\"onwald}

\affiliation[a]{Deutsches Elektronen-Synchrotron DESY,
Platanenallee 6, 15738 Zeuthen, Germany}

\affiliation[b]{
Johannes Kepler University Linz, Research Institute for Symbolic
Computation (RISC),
    Altenberger Stra\ss{}e 69, 4040 Linz, Austria}

\affiliation[c]{Institut f\"ur Theoretische Teilchenphysik,
Karlsruher Institut für Technologie (KIT) D-76128 Karlsruhe, Germany}

\emailAdd{Johannes.Bluemlein@desy.de}
\emailAdd{Peter.Marquard@desy.de}
\emailAdd{Carsten.Schneider@risc.jku.at}
\emailAdd{kay.schoenwald@kit.edu}

\abstract{\noindent 
We report on the calculation of the three--loop polarized and unpolarized flavor non--singlet and 
the polarized singlet anomalous dimensions using massless off--shell operator matrix elements in a 
gauge--variant framework. We also reconsider the unpolarized two--loop singlet anomalous dimensions 
and correct errors in the foregoing literature.}

\FullConference{%
  Loops and Legs in Quantum Field Theory - LL2022,\\
  25-30 April, 2022\\
  Ettal, Germany
}

\begin{document}
\maketitle

\section{Introduction}

\vspace*{1mm}
\noindent
In these proceedings we report on recent results of the calculation of unpolarized and polarized three--loop 
anomalous dimensions based on massless off--shell operator matrix elements (OMEs) 
\cite{Blumlein:2021enk,Blumlein:2021ryt} and the unpolarized two--loop case \cite{Blumlein:2022ndg} correcting 
previous results in Refs.~\cite{Matiounine:1998ky,Matiounine:1998re}.

The unpolarized and polarized non--singlet and singlet anomalous dimensions have been calculated at one--
\cite{ONELOOP}, two-- \cite{TWOLOOP,Blumlein:2022ndg,Hamberg:1991qt}, and three--loop order 
\cite{Moch:2004pa,Vogt:2004mw,Ablinger:2014nga,
Moch:2014sna,Ablinger:2017tan,Behring:2019tus,Blumlein:2021enk,Blumlein:2021ryt,GEHRM}. Here different techniques
as off--shell massless OMEs, the forward Compton amplitude, in part also with scalar and gravitational currents,
and massive on--shell OMEs have been used. In the latter case one obtains the contributions $\propto T_F$, which 
are the complete anomalous dimensions in the cases $(\Delta) \gamma_{qq}^{(2),\rm PS}$ and $(\Delta) 
\gamma_{qg}^{(2)}$.

The method of massless off--shell OMEs is the traditional way to calculate the anomalous dimensions, cf.~
[1a-c]. However, it is a gauge--dependent environment implying in the unpolarized singlet case new operator 
mixings \cite{Hamberg:1991qt,GAUGE,Blumlein:2022ndg}. It is our goal to set up a program chain allowing for a fully 
automated calculation
of the three--loop anomalous dimensions without making structural assumptions motivated by QCD or the 
expected representation of the final results in terms of harmonic sums \cite{Vermaseren:1998uu,Blumlein:1998if}.
The paper is organized as follows.
We discuss first the basic formalism and describe then the calculation method, before we present some examples
for the three--loop anomalous dimensions.
\section{Basic Formalism}

\vspace*{1mm}
\noindent
We form the expectation values of the local twist--2 operators in the unpolarized and polarized case, 
cf. e.g.~\cite{Blumlein:2012bf}, between off--shell quark and gluon states and consider the physical 
projections, which are calculated to three--loop order. There are also other contributions due to the 
violation of the equation of motion (EOM), which, however, are not related to the anomalous dimensions. 
From the physical projection we extract the three--loop anomalous dimensions from the $O(1/\ep)$ terms,
where $\ep = D-4$ denotes the dimensional parameter. The corresponding amplitudes are gauge--variant, 
i.e. they depend on the gauge parameter $\xi$ in the $R_\xi$ gauges. The structure of the pole terms are 
fully predicted by the renormalization group and do depend on lower order expansion coefficients to higher 
orders in $\ep$. In the unpolarized case, the mixing of more local operators has to be considered 
\cite{Hamberg:1991qt,GAUGE,Matiounine:1998ky,Matiounine:1998re,Blumlein:2022ndg}, which we studied up to two--loop order
in \cite{Blumlein:2022ndg}.

In the polarized singlet case the anomalous dimensions are first calculated in the Larin scheme \cite{Larin:1993tq},
which is a consistent scheme. Finally, we transform the anomalous dimensions to the M--scheme 
\cite{Matiounine:1998re,Moch:2014sna}. 

\section{Details of the calculation}
\label{sec:3}

\vspace*{1mm} 
\noindent
The calculation of the unrenormalized massless off--shell OMEs is performed in the following way.
Diagram generation, the performance of Lorentz/Dirac and color algebra are performed by using the 
packages {\tt QGRAF, FORM} and {\tt color} \cite{Nogueira:1991ex,Bierenbaum:2009mv,FORM,
vanRitbergen:1998pn}.
The local operators are resummed into propagators by observing the current crossing relations, 
cf.~\cite{Politzer:1974fr,Blumlein:1996vs}, as has been described in Ref.~\cite{Ablinger:2014yaa},
\begin{eqnarray}
\sum_{N=0}^\infty (\Delta.k)^N \left(t^N \pm (-t)^N\right) \rightarrow
\left[\frac{1}{1 - \Delta.k~t} \pm  \frac{1}{1 + \Delta.k~t}\right].
\label{eq:resu1}
\end{eqnarray}
The three--loop anomalous dimensions are obtained from the contributions of $O(1/\ep)$, determining the 
$N$th moment analytically.

In the individual channels there are up to $O(1600)$ Feynman diagrams. They are reduced to up to $O(250)$ master 
integrals using the code 
{\tt Crusher} \cite{CRUSHER} by applying the integration--by--parts relations \cite{IBP,Chetyrkin:1981qh}.
Coupling constant and wave function renormalizations are performed \cite{BETA,WF} and results from lower order 
factorizing diagrams \cite{Blumlein:2022ndg} are accounted for. The method of arbitrary high Mellin moments
\cite{Blumlein:2017dxp}, implemented 
in the package 
\texttt{SolveCoupledSystem}~\cite{Blumlein:2019hfc}, is used to provide a necessary input set, here of 3000
moments.  The necessary 
initial values for the difference equations
can be obtained from \cite{Chetyrkin:1981qh,INIT}. We then used the method of guessing 
\cite{GUESS,Blumlein:2009tj} and its implementation in {\tt Sage} 
\cite{SAGE,GSAGE} to obtain the recurrences for the different color and
multiple zeta value factors \cite{Blumlein:2009cf}. It turned out that $O(1600)$ moments suffice and the largest
difference equations was of order {\sf o} = {16} and degree  {\sf d} = {304}. These difference equations are 
solved using
methods from difference ring theory \cite{DRING} implemented in the package 
{\tt Sigma} \cite{SIG1,SIG2}. Functions of the package {\tt HarmonicSums} 
\cite{HARMSU,Blumlein:2009ta,Vermaseren:1998uu,Blumlein:1998if,Remiddi:1999ew,
Blumlein:2003gb}
are used to compactify the final results. The automated calculation 
at {\tt Intel(R) Xeon(R) CPU E5-2643 v4} processors amounted to about 40 days 
of CPU time for the projects \cite{Blumlein:2021enk,Blumlein:2021ryt}.

All anomalous dimensions can be expressed in terms of harmonic sums 
\cite{Vermaseren:1998uu,Blumlein:1998if}
\begin{eqnarray}
S_{b,\vec{a}}(N) &=& \sum_{k=1}^N \frac{({\rm sign}(b))^k}{k^{|b|}} S_{\vec{a}}(k),~~~S_\emptyset 
= 
1,~~~b_, a_i \in \mathbb{Z} \backslash \{0\}, N \in \mathbb{N}  \backslash \{0\}. 
\end{eqnarray}
Accordingly, the corresponding $z$-space expressions are given by harmonic polylogarithms ${\rm 
H}_{\vec{a}}(z)$, \cite{Remiddi:1999ew}. Here $z \in [0,1]$ denotes the momentum fraction w.r.t.
the incoming nucleon momentum in the deep--inelastic process.
\section{Anomalous Dimensions}

\vspace*{1mm}
\noindent
We have calculated all the non--singlet anomalous dimensions $\gamma_{qq}^{\pm, \rm NS}, \gamma_{qq}^{s, \rm NS}, 
\Delta \gamma_{qq}^{s, \rm NS}$ for unpolarized and polarized deep-inelastic scattering and also for 
transversity $\gamma_{qq}^{\pm, \rm tr, NS}$, as well as the singlet polarized anomalous dimensions,  
Ref.~\cite{Blumlein:2021enk,Blumlein:2021ryt}. They can all be expressed in terms of harmonic sums, for which
we apply the algebraic relations \cite{Blumlein:2003gb}. If also the structural relations are applied
\cite{Blumlein:2009ta} only 10 harmonic sums contribute. 
\begin{eqnarray}
\{S_1, S_{2,1}, S_{-2,1}, S_{-3,1}, S_{-4,1}, S_{2,1,1}, S_{-2,1,1}, S_{2,1,-2}, S_{-3,1,1}, S_{-2,1,1,1}\}.
\end{eqnarray}

As examples we show one of the transversity anomalous dimensions 
\begin{eqnarray}
&& \hspace*{-7mm} \gamma_{\rm NS}^{(2), \rm tr,+} = \frac{1}{2}\left[1 + (-1)^N\right] \nonumber\\ && \times 
\Biggl\{
\textcolor{blue}{C_F} \Biggl\{
        \textcolor{blue}{T_F^2 N_F^2} \Biggl[
                \frac{8 \big(
                        -8+17 N+17 N^2\big)}{9 N (1+N)}
                -\frac{128}{27} S_1 
                -\frac{640}{27} S_2 
                +\frac{128}{9} S_3 
        \Biggr]
\nonumber\\ && 
        +\textcolor{blue}{C_A T_F N_F} \Biggl[
                -
                \frac{16 \big(
                        -22+45 N+45 N^2\big)}{9 N (1+N)}
                +\Biggl(
                        -\frac{16 \big(
                                9+209 N+209 N^2\big)}{27 N (1+N)}
                        +64 S_3 
\nonumber\\ &&                  
       +\frac{256}{3} S_{-2,1} 
                        -128 \zeta_3
                \Biggr) S_1 
                +\frac{5344}{27} S_2 
                -\frac{448}{3} S_3 
                +\frac{320}{3} S_4 
                +\Biggl(
                        -\frac{1280}{9} S_1 
                        +\frac{128}{3} S_2 
                \Biggr) S_{-2} 
\nonumber\\ &&                
 +\Biggl(
                        -\frac{640}{9}
                        +\frac{128}{3} S_1 
                \Biggr) S_{-3} 
                +\frac{128}{3} S_{-4} 
                -\frac{256}{3} S_{3,1} 
                +\frac{1280}{9} S_{-2,1} 
                +\frac{128}{3} S_{-2,2} 
                -\frac{512}{3} S_{-2,1,1} 
\nonumber\\ &&                
 +96 \zeta_3
        \Biggr]
        +\textcolor{blue}{C_A^2} \Biggl[
                \frac{-968+1657 N+1657 N^2}{18 N (1+N)}
                +\Biggl(
                        \frac{4 P_{14}}{3 (-1+N) N (1+N) (2+N)}
                        -176 S_3 
\nonumber\\ &&                
         -256 S_4 
                        +512 S_{3,1} 
                        -\frac{704}{3} S_{-2,1} 
                        -1024 S_{-2,2} 
                        -1024 S_{-3,1} 
                        +2048 S_{-2,1,1} 
                \Biggr) S_1 
                +\Biggl(
                        -128 S_3 
\nonumber\\ &&                
         -512 S_{-2,1} 
                \Biggr) S_1 ^2
                +\Biggl(
                        -\frac{8344}{27}
                        +384 S_3 
                        +1536 S_{-2,1} 
                \Biggr) S_2 
                +\frac{3112}{9} S_3 
                -\frac{880}{3} S_4 
                +64 S_5 
\nonumber\\ &&                
 +\Biggl(
                        \frac{16 P_2}{(-1+N) N (1+N) (2+N)}
                        +
                        \frac{32 \big(
                                -241+134 N+134 N^2\big) S_1 }{9 (-1+N) (2+N)}
                        -\frac{352}{3} S_2 
                        -64 S_3 
\nonumber\\ &&                
         -1536 S_{2,1} 
                        +128 S_{-2,1} 
                        -192 \zeta_3
                \Biggr) S_{-2} 
                +\Biggl(
                        48
                        -192 S_1 
                \Biggr) S_{-2} ^2
                +\Biggl(
                        256 S_1 ^2
                        -768 S_2 
                        -320 S_{-2} 
\nonumber\\ &&                 
+        \frac{32 \big(
                                -107+67 N+67 N^2\big)}{9 (-1+N) (2+N)}
                        -\frac{352}{3} S_1 
                \Biggr) S_{-3} 
                +\Biggl(
                        -\frac{208}{3}
                        +320 S_1 
                \Biggr) S_{-4} 
                -704 S_{-5} 
                -384 S_{2,3} 
\nonumber\\ &&                
 -768 S_{2,-3} 
                +\frac{704}{3} S_{3,1} 
                +384 S_{4,1} 
                -\frac{64 \big(
                        -107+67 N+67 N^2\big) S_{-2,1} }{9 (-1+N) (2+N)}
                -\frac{352}{3} S_{-2,2} 
\nonumber\\ &&                
 +1088 S_{-2,3} 
                -448 S_{-4,1} 
                +1536 [S_{2,1,-2} 
                + S_{-2,2,1} 
                + S_{-3,1,1}] 
                -768 S_{3,1,1} 
\nonumber\\ &&                 
+\frac{1408}{3} S_{-2,1,1} 
                +512 S_{-2,1,-2} 
                -3072 S_{-2,1,1,1} 
                -\frac{24 \big(
                        -6+5 N+5 N^2\big) \zeta_3}{(-1+N) (2+N)}
        \Biggr]
\Biggr\}
\nonumber\\ && 
+\textcolor{blue}{C_F^2} \Biggl\{
        \textcolor{blue}{T_F N_F} \Biggl[
                92
                +\Biggl(
                        -\frac{8 \big(
                                -8+55 N+55 N^2\big)}{3 N (1+N)}
                        +\frac{1280}{9} S_2 
                        -\frac{512}{3} S_3 
                        -
                        \frac{512}{3} S_{-2,1} 
                        +128 \zeta_3
                \Biggr) 
\nonumber\\ &&  \times
S_1 
                -\frac{80}{3} S_2 
                -\frac{128}{3} S_2 ^2
                +\frac{1856}{9} S_3 
                -\frac{512}{3} S_4 
                +\Biggl(
                        \frac{2560}{9} S_1 
                        -\frac{256}{3} S_2 
                \Biggr) S_{-2} 
                +\Biggl(
                        \frac{1280}{9}
\nonumber\\ &&                
         -\frac{256}{3} S_1 
                \Biggr) S_{-3} 
                -\frac{256}{3} S_{-4} 
                +\frac{256}{3} S_{3,1} 
                -\frac{2560}{9} S_{-2,1} 
                -\frac{256}{3} S_{-2,2} 
                +\frac{1024}{3} S_{-2,1,1} 
                -96 \zeta_3
        \Biggr]
\nonumber\\ && 
        +\textcolor{blue}{C_A} \Biggl[
                -\frac{151}{2}
                +\Biggl(
                        -\frac{8 \big(
                                -206+211 N+211 N^2\big)}{3 (-1+N) N (1+N) (2+N)}
                        -\frac{4288}{9} S_2 
                        +\frac{1984}{3} S_3 
                        +320 S_4 
                        -1024 S_{3,1} 
\nonumber\\ &&                
         +\frac{1984}{3} S_{-2,1} 
                        +3712 S_{-2,2} 
                        +3840 S_{-3,1} 
                        -7168 S_{-2,1,1} 
                \Biggr) S_1 
                +\Biggl(
                        256 S_3 
                        +1792 S_{-2,1} 
                \Biggr) S_1 ^2
\nonumber\\ &&                
 +\Biggl(
                        \frac{604}{3}
                        -832 S_3 
                        -5248 S_{-2,1} 
                \Biggr) S_2 
                +\frac{352}{3} S_2 ^2
                -\frac{6160}{9} S_3 
                +\frac{2416}{3} S_4 
\nonumber\\ &&                
 +\Biggl(
                        -\frac{48 P_2}{(-1+N) N (1+N) (2+N)}
                        +\Biggl(
                                -\frac{64 P_7}{9 (-1+N) N (1+N) (2+N)}
                                -256 S_2 
                        \Biggr) S_1 
\nonumber\\ &&                
         +
                        \frac{992}{3} S_2 
                        +64 S_3 
                        +5376 S_{2,1} 
                        -384 S_{-2,1} 
                        +576 \zeta_3
                \Biggr) S_{-2} 
                +\Biggl(
                        -96
                        +512 S_1 
                \Biggr) S_{-2} ^2
\nonumber\\ &&                
 +\Biggl(
                        -\frac{32 \big(
                                -187+134 N+134 N^2\big)}{9 (-1+N) (2+N)}
                        +\frac{992}{3} S_1 
                        -1152 S_1 ^2
                        +2624 S_2 
                        +960 S_{-2} 
                \Biggr) S_{-3} 
\nonumber\\ &&                
 +\Biggl(
                        \frac{560}{3}
                        -1472 S_1 
                \Biggr) S_{-4} 
                +2304 S_{-5} 
                +768 S_{2,3} 
                +2688 S_{2,-3} 
                -\frac{1856}{3} S_{3,1} 
                -768 S_{4,1} 
\nonumber\\ && 
                +\frac{64 \big(
                        -187+134 N+134 N^2\big) S_{-2,1} }{9 (-1+N) (2+N)}
                +\frac{992}{3} S_{-2,2} 
                -3648 S_{-2,3} 
                +1728 S_{-4,1} 
\nonumber\\ &&                 
-5376 [S_{2,1,-2} 
                + S_{-2,2,1} 
                + S_{-3,1,1}] 
                +1536 [S_{3,1,1} 
                - S_{-2,1,-2} ]
                -\frac{3968}{3} S_{-2,1,1} 
\nonumber\\ &&                 
+10752 S_{-2,1,1,1} 
                +\frac{72 \big(
                        -6+5 N+5 N^2\big) \zeta_3}{(-1+N) (2+N)}
        \Biggr]
\Biggr\}
\nonumber\\ && 
+\textcolor{blue}{C_F^3} \Biggl\{
        -29
        +\Biggl(
                \frac{384 \big(
                        -1+N+N^2\big)}{(-1+N) N (1+N) (2+N)}
                +128 S_2 ^2
                -384 S_3 
                +128 S_4 
                +512 S_{3,1} 
\nonumber\\ &&                
 -384 S_{-2,1} 
                -3328 S_{-2,2}
                                -3584 S_{-3,1} 
                +6144 S_{-2,1,1} 
        \Biggr) S_1 
        -256 S_{-2} ^2 S_1 
        +\Biggl(
                12
                +512 S_3 
\nonumber\\ &&                
 +4352 S_{-2,1} 
        \Biggr) S_2 
        -96 S_2 ^2
        +104 S_3 
        -480 S_4 
        +\Biggl(
                \frac{32 P_2}{(-1+N) N (1+N) (2+N)}
\nonumber\\ &&                
 +\Biggl(
                        \frac{384}{(-1+N) N (1+N) (2+N)}
                        +512 S_2 
                \Biggr) S_1 
                -192 S_2 
                +128 S_3 
                -4608 S_{2,1} 
                +256 S_{-2,1} 
\nonumber\\ &&                
 -384 \zeta_3
        \Biggr) S_{-2} 
        +\Biggl(
                \frac{192}{(-1+N) (2+N)}
                -192 S_1 
                +1280 S_1 ^2
                -2176 S_2 
                -640 S_{-2} 
        \Biggr) S_{-3} 
\nonumber\\ &&         
+\Biggl(
                -96
                +1664 S_1 
        \Biggr) S_{-4} 
        -1792 S_{-5} 
        +384 [-S_{2,3} 
        + S_{3,1} 
        + S_{4,1}] 
        -2304 S_{2,-3} 
\nonumber\\ &&        
-\frac{384 S_{-2,1} }{(-1+N) (2+N)}
        -1536 S_1 ^2 S_{-2,1} 
        -192 S_{-2,2} 
        +2944 S_{-2,3} 
        -1664 S_{-4,1} 
        +4608 S_{2,1,-2} 
\nonumber\\ &&         
-768 S_{3,1,1} 
        +768 S_{-2,1,1} 
        +1024 S_{-2,1,-2} 
        +4608 [S_{-2,2,1} 
        + S_{-3,1,1}] 
        -9216 S_{-2,1,1,1} 
\nonumber\\ &&         
-\frac{48 \big(
                -6+5 N+5 N^2\big) \zeta_3}{(-1+N) (2+N)}
\Biggr\}
\Biggr\}

\end{eqnarray}

and one of the polarized singlet anomalous dimensions,
\begin{eqnarray}
\label{eq:gg2}
\Delta \gamma_{gg}^{(2)} &=&
\textcolor{blue}{C_A T_F^2 N_F^2} \Biggl[
        -\frac{16 P_8}{27 N^2 (1+N)^2} S_1
        -\frac{4 P_{48}}{27 N^3 (1+N)^3}
\Biggr]
+\textcolor{blue}{C_F} \Biggl[
        \textcolor{blue}{T_F^2 N_F^2} \Biggl[
                -\frac{8 P_{59}}{27 N^4 (1+N)^4}
\nonumber\\ &&                
+\frac{64 (N-1) (2+N) \big(
                        -6-8 N+N^2\big)}{9 N^3 (1+N)^3} S_1
                +\frac{32 (N-1) (2+N)}{3 N^2 (1+N)^2} S_1^2
\nonumber\\ && 
                -\frac{32 (N-1) (2+N)}{N^2 (1+N)^2} S_2
        \Biggr]
        +\textcolor{blue}{C_A T_F N_F} \Biggl[
                \frac{8 P_6}{N^3 (1+N)^3} S_2
                -\frac{8 P_9}{3 N^3 (1+N)^3} S_1^2
\nonumber\\ &&  
                +
                \frac{2 P_{77}}{27 (N-1) N^5 (1+N)^5 (2+N)}
                +\Biggl(
                        -\frac{8 P_{67}}{9 (-1+N) N^4 (1+N)^4 (2+N)}
\nonumber\\ &&                        
-\frac{32 (N-1) (2+N)}{N^2 (1+N)^2} S_2
                        +128 \zeta_3
                \Biggr) S_1
                +\frac{32 (N-1) (2+N)}{3 N^2 (1+N)^2} S_1^3
                -\frac{32 \big(
                        34+N+N^2\big)}{3 N^2 (1+N)^2} 
\nonumber\\ && \times
S_3
                +\Biggl(
                        \frac{128P_2}{(N-1) N^2 (1+N)^2 (2+N)} S_1
                        -\frac{32 P_{23}}{(N-1) N^2 (1+N)^3 (2+N)}
                \Biggr) S_{-2}
\nonumber\\ &&                
 -\frac{192 \big(
                        4-N-N^2\big)}{N^2 (1+N)^2} S_{-3}
                +\frac{64 (N-1) (2+N)}{N^2 (1+N)^2} S_{2,1}
                -\frac{128 \big(
                        -8+N+N^2\big)}{N^2 (1+N)^2} S_{-2,1}
\nonumber\\ &&               
  -\frac{64 (-3+N) (4+N)}{N^2 (1+N)^2} \zeta_3
        \Biggr]
\Biggr]
+\textcolor{blue}{C_A^3} \Biggl[
        \frac{64 P_{16}}{9 N^2 (1+N)^2} S_{-2,1}
        -\frac{32 P_{18}}{9 N^2 (1+N)^2} S_3
\nonumber\\ &&         
+\frac{P_{74}}{27 (N-1) N^5 (1+N)^5 (2+N)}
        +\Biggl(
                 \frac{4 P_{69}}{9 (N-1) N^4 (1+N)^4 (2+N)}
\nonumber\\ && 
                -\frac{64 P_{17}}{9 N^2 (1+N)^2} S_2
                +128 S_2^2
                +\frac{16 \big(
                        -96+11 N+11 N^2\big)}{3 N (1+N)} S_3
                +192 S_4
\nonumber\\ &&                
 +\frac{1024}{N (1+N)} S_{-2,1}
                -640 S_{-2,2}
                -768 S_{-3,1}
                +1024 S_{-2,1,1}
        \Biggr) S_1
\nonumber\\ &&         
+\Biggl(
                -\frac{256 \big(
                        1+3 N+3 N^2\big)}{N^3 (1+N)^3}
                +128 S_3
                -256 S_{-2,1}
        \Biggr) S_1^2
        +\Biggl(
                -
                \frac{16 P_{41}}{9 N^3 (1+N)^3}
\nonumber\\ &&                
 +64 S_3
                +640 S_{-2,1}
        \Biggr) S_2
        -\frac{256}{N (1+N)} S_2^2
        -\frac{384}{N (1+N)} S_4
        +64 S_5
\nonumber\\ &&         
+\Biggl(
                \frac{32 P_{52}}{9 (N-1) N^3 (1+N)^3 (2+N)}
                +\big(
                        -\frac{64 P_{32}}{9 (-1+N) N (1+N)^2 (2+N)}
                        +256 S_2
                \Biggr) 
\nonumber\\ &&  \times
S_1
                -\frac{512}{N (1+N)} S_2
                +128 S_3
                -768 S_{2,1}
        \Biggr) S_{-2}
        +\Biggl(
                -\frac{16 \big(
                        24+11 N+11 N^2\big)}{3 N (1+N)}
\nonumber\\ &&                 
+64 S_1
        \Biggr) S_{-2}^2
        +\Biggl(
                -\frac{32 P_{15}}{9 N^2 (1+N)^2}
                -\frac{1536}{N (1+N)} S_1
                +384 S_1^2
                -320 S_2
        \Biggr) S_{-3}
\nonumber\\ &&         
+\Biggl(
                -\frac{1024}{N (1+N)}
                +512 S_1
        \Biggr) S_{-4}
        -192 S_{-5}
        -384 S_{2,-3}
        +\frac{1280}{N (1+N)} S_{-2,2}
\nonumber\\ &&         
+384 S_{-2,3}
        +\frac{1536}{N (1+N)} S_{-3,1}
        -384 S_{-4,1}
        +768 S_{2,1,-2}
        -\frac{2048}{N (1+N)} S_{-2,1,1}
\nonumber\\ &&         
+768 [S_{-2,2,1} + S_{-3,1,1}]
        -1536 S_{-2,1,1,1}
\Biggr]
\nonumber\\ &&
+\textcolor{blue}{C_F^2 T_F N_F} \Biggl[
        -\frac{4 P_{75}}{(N-1) N^5 (1+N)^5 (2+N)}
        +\Biggl(
                \frac{32 (N-1) (2+N) S_2}{N^2 (1+N)^2}
\nonumber\\ &&                 
-\frac{16 P_{42}}{N^4 (1+N)^4}
        \Biggr) S_1
        +\frac{8 (N-1) (2+N) \big(
                2+3 N+3 N^2\big)}{N^3 (1+N)^3} S_1^2
        -\frac{32 (N-1) (2+N)}{3 N^2 (1+N)^2} 
\nonumber\\ && \times S_1^3         
-\frac{8 (2+N) \big(
                2-11 N-16 N^2+9 N^3\big)}{N^3 (1+N)^3} S_2
        +\frac{32 \big(
                10+7 N+7 N^2\big)}{3 N^2 (1+N)^2} S_3
\nonumber\\ &&         
+\Biggl(
                -\frac{64 \big(
                        10+N+N^2\big)}{(N-1) N (1+N) (2+N)}
                +\frac{512}{N^2 (1+N)^2} S_1
        \Biggr) S_{-2}
        +\frac{256}{N^2 (1+N)^2} S_{-3}
\nonumber\\ &&         
-\frac{64 (N-1) (2+N)}{N^2 (1+N)^2} S_{2,1}
        -\frac{512}{N^2 (1+N)^2} S_{-2,1}
        +\frac{192 \big(
                -2-N-N^2\big)}{N^2 (1+N)^2} \zeta_3
\Biggr]
\nonumber\\ && 
+\textcolor{blue}{C_A^2 T_F N_F} \Biggl[
        \frac{32 P_4}{9 N^2 (1+N)^2} S_2
        +\frac{32 P_{11}}{9 N^2 (1+N)^2} S_{-3}
        -\frac{64 P_{11}}{9 N^2 (1+N)^2} S_{-2,1}
\nonumber\\ &&         
+\frac{16 P_{13}}{9 N^2 (1+N)^2} S_3
        +\frac{2 P_{76}}{27 (N-1) N^5 (1+N)^5 (2+N)}
        +\Biggl(
                \frac{1280}{9} S_2
                -\frac{64}{3} S_3
\nonumber\\ && 
                -\frac{8 P_{68}}{27 (-1+N) N^4 (1+N)^4 (2+N)}
                -128 \zeta_3
        \Biggr) S_1
        +\frac{64}{3} S_{-2}^2
\nonumber\\ &&         +\Biggl(
                \frac{64 P_{45}}{9 (N-1) N^2 (1+N)^2 (2+N)} 
S_1
                -\frac{32 P_{50}}{9 (N-1) N^3 (1+N)^3 (2+N)}
        \Biggr) S_{-2}
\nonumber\\ &&
        +\frac{128 \big(
                -3+2 N+2 N^2\big)}{N^2 (1+N)^2} \zeta_3
\Biggr].
\end{eqnarray}


The polynomials are given in Refs.~\cite{Blumlein:2021enk,Blumlein:2021ryt}. The small $z$ and large $z$
limits of the anomalous dimensions have also been considered in explicit form. The latter are related to the cusp 
anomalous dimensions. There are various partial predictions on the small $z$ behaviour, however, not specifying the 
factorization scheme used. These are theoretically interesting, but are not of quantitative importance since it 
is known for long that subleading terms do strongly modify the leading order effects over about three additional 
subleading  orders \cite{SX}.

We have also recalculated the unpolarized two--loop anomalous dimensions using the method of massless 
off--shell OMEs, \cite{Blumlein:2022ndg}. Here new local operators contribute \cite{Hamberg:1991qt,GAUGE,Blumlein:2022ndg} 
to cancel the 
gauge--variant 
contributions to
obtain the anomalous dimensions. We corrected a series of errors contained in 
\cite{Matiounine:1998ky,Matiounine:1998re} and provided all expansion coefficients emerging at two--loop order
\cite{Blumlein:2022ndg}, which are relevant for upcoming four--loop calculations.
\section{Conclusions}

\vspace*{1mm}
\noindent
We have calculated the three--loop polarized and unpolarized non--singlet anomalous dimensions, 
including transversity, and as well the polarized three--loop singlet anomalous dimensions applying
the traditional method of massless off--shell operator matrix elements, a gauge--variant method,
which requires special projections. The calculations have been performed without assuming special 
conditions concerning the structure of the final result. The method of arbitrary high Mellin moments 
plays a central role in these computations in establishing the corresponding difference equations 
through the method of guessing, after having exploited the equations for the master integrals. These 
equations are solved subsequently using algorithms of difference ring theory. We agree with all the 
results previously obtained in the literature. The polarized non--singlet anomalous dimension 
$\Delta \gamma_{qq}^{s, \rm NS}$, related to the polarized structure function $g_5^{-}$, \cite{Blumlein:1996vs}, 
has been calculated using the associated forward Compton amplitude. In the unpolarized singlet case also new OMEs 
contribute. Here we have calculated all contributions emerging at the two--loop level and corrected results 
in the literature. The present method is suited to be expanded to the four--loop level. The knowledge of the 
higher--loop anomalous dimensions form one important asset for the precise description of the scaling violations 
of the deep--inelastic structure functions, which provide an important way to measure the QCD coupling constant 
$\alpha_s(M_Z^2)$ \cite{ALPHAS} at highest precision possible.


\end{document}